# Comparative Analysis of Polynomial and Rational Approximations of Hyperbolic Tangent Function for VLSI Implementation

Mahesh Chandra[1]

[1] NXP Semiconductors, India
Mahesh.chandra_1@nxp.com

**Abstract.** Deep neural networks yield the state-of-the-art results in many computer vision and human machine interface applications such as object detection, speech recognition etc. Since, these networks are computationally expensive, customized accelerators are designed for achieving the required performance at lower cost and power. One of the key building blocks of these neural networks is non-linear activation function such as sigmoid, hyperbolic tangent (tanh), and ReLU. A low complexity accurate hardware implementation of the activation function is required to meet the performance and area targets of the neural network accelerators. Even though, various methods and implementations of tanh activation function have been published, a comparative study is missing. This paper presents comparative analysis of polynomial and rational methods and their hardware implementation

## 1 Introduction

An artificial neuron, modelled around the biological neurons, consists of a MAC (multiply and accumulate) functional unit and a non-linear activation unit. Artificial neural networks (ANNs), developed using the interconnection of such neurons, are used for machine learning application as they can learn complex relationships between the inputs and outputs. These neural networks have yielded state of the art results in various applications. It is proven now that deeper neural networks have better learning capabilities. Since, these algorithms require huge computing resources; dedicated accelerators are being developed to speed up the execution. Activation function is one of the key building blocks required for the efficient hardware accelerator.

Activation functions used in ANNs have evolved over time. Till a few years back, sigmoid and hyperbolic tangent (tanh) were used more frequently; however, simpler activation functions such as rectified linear unit (ReLU) have been more popular in recent deep neural networks. Most feed forward neural networks prefer ReLU over other activation function for its various useful properties such as lower compute complexity, easy trainability etc. However, some applications require sequence modelling and use recurrent neural networks (RNNs) and long short-term memory (LSTM) topologies. Tanh is still an integral part of these neural networks [1,2].

Tanh function, shown in figure 1, is a non-linear function defined as:

$$tanh(x) = \frac{e^x - e^{-x}}{e^x + e^{-x}} \tag{1}$$

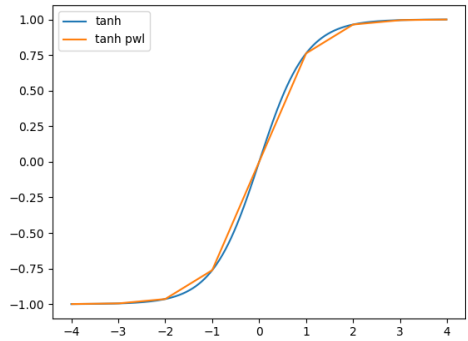

Fig. 1. tanh function and its piecewise linear approximation

The simplest implementation is to store the values of the function in a lookup table (LUT) and approximate the output with the lookup table value for the nearest input. Since, the function is non-uniform, it's challenging to balance the tradeoff between accuracy and area if the range is divided in equal sub-ranges. To address this issue, Leboeuf et al. used range addressable lookup table [3], and Namin et al. used a two-step LUT; one for coarse estimation and another for the finer estimation [4]. Tanh is an S-shaped odd function and can be divided into saturation and processing regions. Zamanlooy et al. designed the hardware by dividing it in three regions and optimizing the design specific to each of them [5].

Interpolation can be used to reduce the error in function approximation. Piecewise linear (PWL) approximation is one of the most popular method for implementing the tanh activation function [1, 2, 6]. Taylor series expansion has been used to reduce the error by Adnan et al. [7]. Other approaches for interpolation such as the DCT (discrete cosine transform) interpolation filter (DCTIF) have also been used for tanh approximation in non-saturation region [8].

Rational approximations such as Padé Approximant [9] and Lambert's continuous fraction [10] have also been explored by the researchers. Trigonometric identity for tanh of sum of two angles is used by Langhammer and Pasca to compute the tanh value along with other approximations [11]. Ron Doerfler has published an interesting method for fast oral calculation of various trigonometric and exponential function [12]. This could also be an interesting method for hardware implementation.

Chen at al. have used the CORDIC method for implementing hyperbolic tangent function. They use two CORDIC functions; one for computing sinh and cosh, and another one for divider [13]. Gomar et al. approximate the tanh function by another simpler exponential function of base two for hardware implementation [14]. Their implementation requires an exponential unit, a division unit and supporting logic.

In hardware accelerator designs, hyperbolic tangent (tanh) function, being a non-linear function, requires specific consideration for the accuracy and area trade-off. It is evident from this short list that various approximations and corresponding hardware implementations have been published by the research community. Some of them are too complex and require huge resources and may be overkill for applications which work with fixed point data such as deep learning. For such applications, there is no comparative analysis of relatively simpler methods of polynomial and rational approximations. Though, some of these methods are reviewed by some researchers [15], a comprehensive review of these methods is missing. This paper tries to fill this gap by discussing different polynomial and rational methods and their hardware implementations.

## 2 Discussion on approximations

Following approximations have been considered for the analysis in this paper.

### 2.1 Piecewise Linear Interpolation (PWL)

PWL is one the simplest and most popular method of approximation. The function is divided into multiple linear partitions and the function values at the endpoints of these partitions are stored in an LUT. These are then used for approximating the function linearly in the range. Given the function values at $a$ and $b$; the approximate value of the function at $a<x<b$ is approximated by:

$$\widehat{f(x)} = f(a) + \frac{f(b)-f(a)}{b-a} \times (x-a) \tag{2}$$

The domain may be divided uniformly or non-uniformly. The uniform division simplifies the implementation while the non-uniform division reduces storage requirement. Algorithms are available for selecting most significant points given error tolerance. Selection of these points and precision affect the accuracy of the approximation.

### 2.2 Taylor Series Expansion

Taylor series expansion for a function $f(x)$ is given by:

$$f(x) = \sum_{n=0}^{\infty} \frac{f^{(n)}(h)}{n!}(x-h)^n \tag{3}$$

The function can be approximated by the sum of first $K$ terms, as given in (4), depending on required accuracy.

$$\widehat{f(x)} = f(h) + f'(h)(x-h) + \frac{f''(h)}{2!}(x-h)^2 + .. + \frac{f^{(K-1)}(h)}{K!}(x-h)^{K-1} \tag{4}$$

Number of terms used for approximation (i.e. $K$) and distance of $x$ from $h$ affects the accuracy. To improve the accuracy with smaller number of terms (trade-off between the logic and storage area), the function and its derivatives can be stored at multiple points in the domain of $f(x)$. The derivatives of the tanh function have an excellent property that they are also a function of tanh. This property can be exploited to reduce the storage requirement by storing only the function values and computing the derivatives on run-time. The derivatives are given by:

$$f'(x) = 1 - tanh^2(x) = 1 - f^2(x) \tag{5}$$

$$f''(x) = -2f(x) \times f'(x) = 2[f^3(x) - f(x)] \tag{6}$$

$$f'''(x) = -2[f(x) \times f''(x) + \{f'(x)\}^2] = -2[1 - 4f^2(x) + 3f^4(x)] \tag{7}$$

### 2.3 Catmull-Rom Spline Interpolation

Splines are used for generating the curves of various shapes in the computer graphics applications. The approximation spline functions do not pass through the interpolating points while the interpolation spline function pass through them. Catmull-Rom spline function [17] is an interpolating spline, which has been used in many graphics and engineering applications. Given control points, the shape of the curve is fixed for the Catmull-Rom spline functions; however, there are variations of this function which can be adapted to a given shape [18]. Moreover, a cubic Catmull-Rom spline function has only integer coefficients which reduces the implementation cost when compared to other spline functions. These properties of Catmull-Rom spline are very useful for the hardware implementation of tanh function.

The uniform cubic Catmull-Rom spline is defined as:

$$\widetilde{f(x)} = \tfrac{1}{2} \begin{bmatrix} t^3 & t^2 & t & 1 \end{bmatrix} \begin{bmatrix} -1 & 3 & -3 & 1 \\ 2 & -5 & 4 & -1 \\ -1 & 0 & 1 & 0 \\ 0 & 2 & 0 & 0 \end{bmatrix} \begin{bmatrix} P_{k-1} \\ P_k \\ P_{k+1} \\ P_{k+2} \end{bmatrix} \tag{8}$$

Where,
$P_i$ is value of function (e.g. tanh) at the uniformly sampled $x_i$ in given input range,
$t$ is between 0 and 1 and used to compute interpolation factor, and
$\widetilde{f(x)}$ is the interpolated value of the function at $x$ for $x_k < x < x_{k+1}$
Number of control points used for interpolation affects the accuracy.

### 2.4 Trigonometric Expansion using Velocity Factor Method

Hyperbolic tangent function can be approximated very fast using two simple algebraic manipulations given the function value at some points [12]. Hyperbolic tangent for the addition of two angles is given by:

$$\tanh(a + b) = \frac{\tanh a + \tanh b}{1 + \tanh a \times \tanh b} \tag{9}$$

Given tanh value at $a$, and very small $b$, it can be approximated as below:

$$\tanh b = b$$

$$\tanh(a + b) = \frac{\tanh a + b}{1 + b \times \tanh a}$$

$$\tanh(a + b) = (\tanh a + b) \times (1 - b \times \tanh a)$$

$$\tanh(a + b) = \tanh a + b \times (1 - \tanh^2 a) \tag{10}$$

This approximation works well for small '$b$'. For large differences, we can directly use equation (9). However, it requires operations which are costly for hardware or software implementation and difficult to parallelize. An alternative representation, that simplifies these operations, is presented in [12] and reproduced below.

Instead of storing tanh values in the LUTs, store velocity factor $f$, defined as:

$$f_a = \frac{1+\tanh a}{1-\tanh a} \qquad (11)$$

To compute tanh from *f*, we can use following equation.

$$\tanh a = \frac{f_a - 1}{f_a + 1} \qquad (12)$$

Given $f_a$ and $f_b$, $f_{a+b}$ can be computed as:

$$f_{a+b} = \frac{1 + \tanh(a+b)}{1 - \tanh(a+b)}$$

$$f_{a+b} = \frac{(1 + \tanh a) \times (1 + \tanh b)}{(1 - \tanh a) \times (1 - \tanh b)}$$

$$f_{a+b} = f_a \times f_b \qquad (13)$$

For implementation, velocity factor can be stored in an registers (or LUTs) for the numbers which are the power of two and more than a threshold (e.g. 2⁻⁶). Equations (12) and (13) are used to compute the tanh value for the sum of stored numbers. For the addition smaller than threshold, equation (10) is used for compensating the error.

The smallest resolution, for which velocity factor is stored in the registers, affects the accuracy of the approximation.

### 2.5   Lambert's continuous fraction

Hyperbolic tangent unction can be written using Lambert's continuous fraction expansion as:

$$\tanh(x) = \cfrac{x}{1 + \cfrac{x^2}{3 + \cfrac{x^2}{5 + \cfrac{x^2}{7 + \dots}}}} \qquad (14)$$

The function can be approximated by using just first few divisions. For *K* division terms, the function approximation can be written in a nice iterative way as given below [19]:

$$T_{-1} = 1; \; T_0 = 2 \times K + 1$$

$$T_n = (2 \times K + 1 - 2 \times n) \times T_{n-1} + x^2 \times T_{n-2} \qquad \forall 1 \le n \le K$$

$$\widetilde{f(x)} = \frac{x \times T_{K-1}}{T_K} \qquad (15)$$

The number of division terms (i.e. *K*) affects the accuracy of the approximation.

## 3   Error Analysis

In case of the deep learning applications, it has been shown that the inference is less sensitive to the quantization of the data (i.e. the precision). For the purpose of this analysis, we can assume the data to be represented as 16-bit or 8-bit fixed point signed input to the tanh block.

### 3.1 Domain for function approximation

For 16-bit fixed point input data, we can consider 13-bit or 12-bit precision for fractional part. The range of the input data in this case will be either (-4,4) or (-8,8) respectively.

For practical purposes, we can constrain the domain to $\tanh^{-1}[\pm(1-2^{-b})]$. For 8, 12 and 16-bit signed fixed-point representation with fractional only (fractional with one-bit integer), the corresponding domain is ±2.77 (±2.42), ±4.16 (±3.82) and ±5.55 (±5.20) respectively. So, we can constrain the domain to (-6,6) for analysis and output the maximum value (i.e. $\pm (1-2^{-b})$) beyond this domain. In such a case, the error for tanh is smaller than that can be represented by the least significant bit.

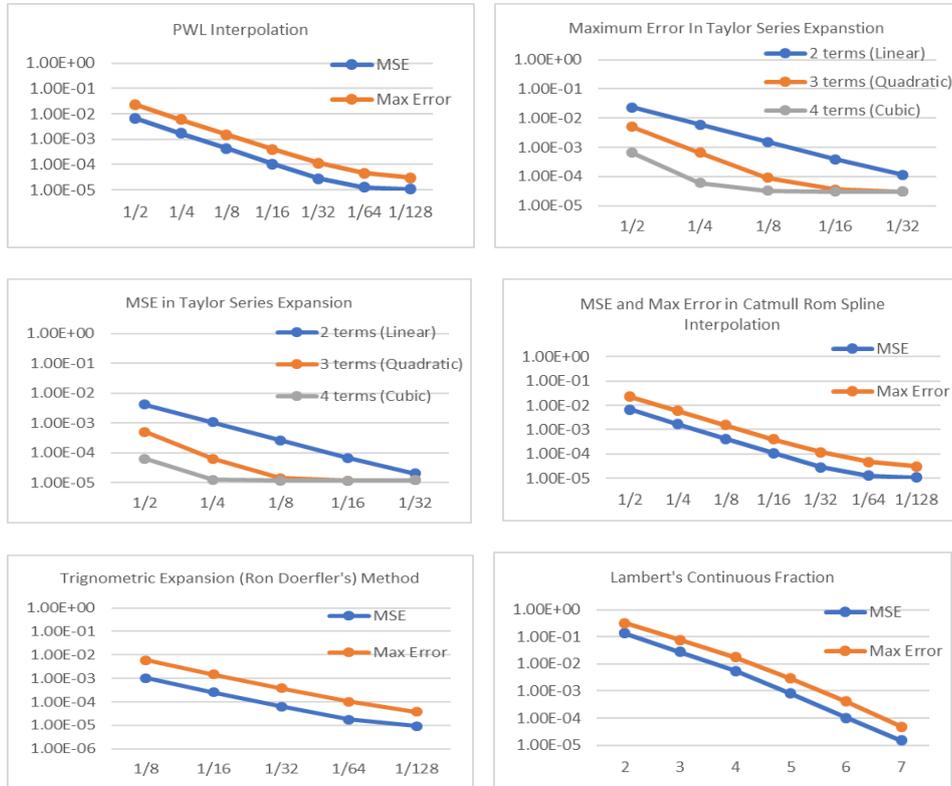

Fig. 2 Maximum absolute and mean square error (MSE) as a function of configuration parameter for various approximations

### 3.2 Maximum Error

For this analysis, maximum error is restricted to one bit (i.e. $1 ulp$).

### 3.3 Method of Analysis

To analyze the error for different approximations, the code was written in python and the maximum absolute error and mean square error (MSE) is computed for differ-ent configurations. Implementation of tanh used in numpy python library [16] is used as reference for error analysis. After this, the parameters are selected for different approximations for the hardware complexity analysis. These parameters are chosen such that the error for different approximation is in similar range.

### 3.4 Error Plots

Figure 2 presents the error analysis for different approximations with respect to the tunable parameters. The Y-axis shows the maximum (absolute) error and mean square error (MSE) and X-axis shows the configuration parameter which is specific to the approximation.

Accuracy of PWL depends on the distance between the interpolated points (step size) which is shown on the X-axis. Taylor series approximates the function very well around a point at which the value of function and its derivatives is known. So, there are two parameters to choose, number of terms and number of points at which the function value is stored. These points are chosen uniformly and presented as step size in X-axis. Accuracy of Catmull Rom interpolation can be increased by increasing number of control points. For uniform Catmull Rom interpolation, the step size, shown in X-axis, is the control parameter

In the case of trigonometric expansion using velocity factor method, the accuracy is governed by the threshold below which the tanh is approximated as linear function. The chart below plots the maximum error and MSE (Y-axis) as a function of this threshold (X-axis). Accuracy of approximation using continuous fraction method depends on the number of continuous fractions used for computation, shown on X-axis.

## 4 Design Complexity Analysis

This section presents the analysis of the design complexity for the digital hardware implementation of the tanh function. Since, tanh is an odd function, the main algorithm can be implemented for positive values only.

### 4.1 Configurations selected for analysis

**Table 1** Configurations selected for analysis (max input 6.0, 12 bit input precision, 15 bit output precision)

| Approximation Method | Step Size | No. of Terms | MSE | Max Error |
|---|---|---|---|---|
| **PWL (A)** | 1/64 | NA | $1.24 \times 10^{-5}$ | $4.65 \times 10^{-5}$ |
| **Taylor 1 (B1)** | 1/16 | 3 (Quadratic) | $1.16 \times 10^{-5}$ | $3.65 \times 10^{-5}$ |
| **Taylor 2 (B2)** | 1/8 | 4 (Cubic) | $1.17 \times 10^{-5}$ | $3.23 \times 10^{-5}$ |
| **Catmull Rom (C)** | 1/16 | NA | $1.13 \times 10^{-5}$ | $3.63 \times 10^{-5}$ |
| **Trig Expansion (D)** | 1/128 | NA | $9.53 \times 10^{-6}$ | $3.85 \times 10^{-5}$ |
| **Lambert (E)** | NA | 7 | $1.50 \times 10^{-5}$ | $4.87 \times 10^{-5}$ |

The accuracy of the approximations depends on the configuration parameters such as step size, number of terms. For a fair comparison, the configuration must be chosen such that their performance is similar. For this purpose, the analysis has been performed for 16-bit signed input x such that -6 < x < 6. The precision of fractional part for the input is determined as 2-12 from this range as 12-bits are left for fractional part after 1 bit for sign and 3 for the integer part. Output is assumed signed fractional number of 16-bits. This fixes the precision to be to be 2-15 for both the output and LUT entries.

Based on the analysis, configurations shown in table 1 are selected. Note that to keep the table compact, the threshold in case of trigonometric expansion using velocity factor method (D) is shown under the "Step Size" column and the number of fractions in Lambert's case is shown under "number of terms" column.

### 4.2    Piecewise Linear Approximation (PWL)

PWL function can be implemented in hardware by LUTs and interpolation logic. Divider is not required as the denominator (b-a) in (2) is fixed for a given step size. The most significant bits of the input are used as address to the LUT and the remaining least significant bits are used for interpolation as shown in figure 3.

Since, the stored values are fixed (i.e. constant), we can use bitmapping (combinatorial) logic instead of a memory cut to store these values for smaller area. Since, the LUT are hardwired and two locations need to be fetched for interpolation; the LUT is split in two with alternate entries to save latency. Such a digital logic implementation requires two adders, one multiplier and two LUTs with 384 (128×6/2) entries each.

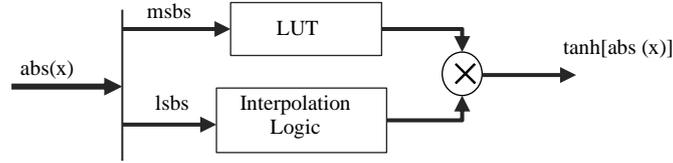

Fig.3 High level Block diagram for polynomial approximation methods  (A, B1, B2 and C)

To improve the performance or precision, the LUT size may require changes as more values may be needed. Increasing LUT size results in reduced operating frequency. So, scaling the solution is not very easy.

### 4.3    Taylor Series Expansion

As discussed earlier, there are two parameters to choose, number of terms and number of points at which the function value is stored. Two different configurations B1 and B2 are chosen; one with smaller LUT size and another with smaller polynomial degree. For logic, each degree of polynomial requires one adder and one multiplier as illustrated below.

$$a_0 + a_1 x + a_2 x^2 + .. + a_{n-1} x^{n-1} + a_n x^n = a_0 + x(a_1 + x(a_2 + ..x(a_{n-1} + a_n x))..)) \quad (16)$$

The hardware can be implemented using two adders, two multipliers and an LUT of 96 entries, or three adders, three multipliers and an LUT of 48 entries for these configurations. Addressing of LUTs and interpolation factor scheme remains same as discussed above for PWL. With regards to the coefficients $a_i$, these can be computed on runtime using tanh values as given in (5)-(7) or can be stored in LUTs.

To improve the performance or precision, the LUT size may require changes. Increasing LUT size results in reduced operating frequency. So, scaling the solution is not very easy; however, it is better than PWL because of smaller LUT size.

### 4.4 Catmull Rom Spline Interpolation

The cubic Catmull-Rom spline given in (8) can be re-written as (17).

$$f(x) = [P_{k-1} \quad P_k \quad P_{k+1} \quad P_{k+2}] \begin{bmatrix} -t^3 + 2t^2 - t \\ 3t^3 - 5t^2 + 2 \\ -3t^3 + 4t^2 + t \\ t^3 - t^2 \end{bmatrix} \tag{17}$$

The equation (17) can be considered as a dot product of two vectors and can be implemented by a simple MAC and vector computation units [20]. The first vector, P vector, contains the control points, and the second vector, t vector, contains the interpolation factor. The control points used in P-vector are stored in a look up table. The most significant five bits of the input are used as the index to the LUT. The second vector, t vector, contains the interpolation factor and can either be computed by a digital circuit or can be stored in the LUT depending on performance area tradeoff. The LUT implementation can be operated at higher frequency, while the polynomial computation logic is smaller in area. The digital logic needs to compute the values of the four cubic polynomials of t. As discussed in previous paragraph, msbs are used for addressing the LUT, the remaining bits (lsbs) can directly be used as t.

### 4.5 Trignometric Expansion using Velocity Factor Method

The basic implementation of velocity factor computation requires 10 entry LUTs storing the tanh velocity factor (VF) value for 2k (-7 ≤ k ≤ 2), and 9 multipliers (one for each bit). Note that the LUT is implemented by multiplexers as shown in fig 4.

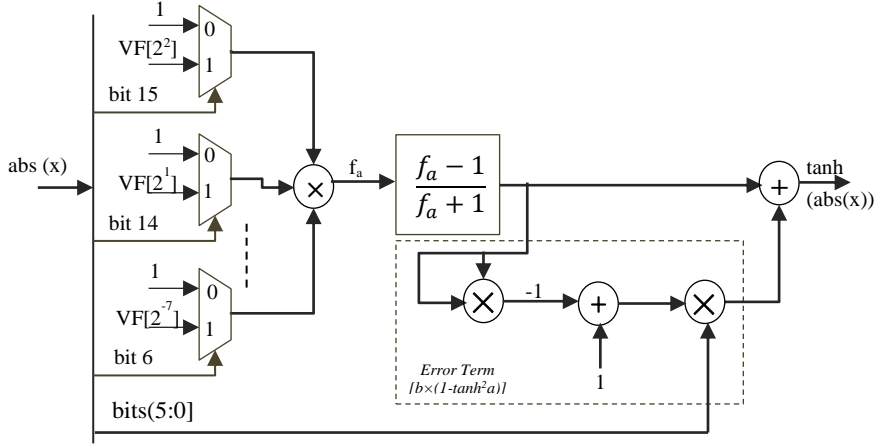

Fig. 4. High level Block diagram for trignometric expansion method (D)

However, simple optimization can be used to reduce number of multipliers at the cost of the LUT size. Instead of storing values of tanh velocity factor values for a single bit, we can store the velocity factor of combination of bits. For example, we can combine two bits and store the values as given by table 2.

**Table 2.** Multi-bit lookup for velocity factors

| bits | Velocity factor |
|------|-----------------|
| **00** | 1.0 |
| **01** | Velocity factor corresponding to lsb |
| **10** | Velocity factor corresponding to msb |
| **11** | Multiplication of velocity factors of lsb and msb |

This scheme requires 20 LUT entries and 4 multipliers (for 1/256 threshold). The LUT is implemented by multiplexers and instead 2-to-1 multiplexers, 4-to-1 multiplexers are used. Apart from these, we need two adders, and a divider for coarse tanh approximation. Further refinement requires two adders, one multiplier and a square unit. Division can be performed by multiplying numerator with the reciprocal of denominator which can be computed using Newton Raphson method [21]. More optimizations can be done for implementing the method [22].

### 4.6 Lambert's continuous fraction

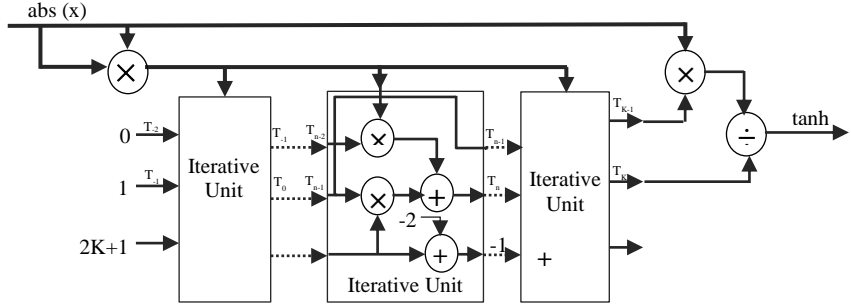

Fig. 5. High level block diagram of iterative continuous fraction method (E)

The iterative method provides a nice pipelined implementation. It requires two adders and two multipliers in each stage except the first two. Number of stages is equal to the number of fractions. The last step requires one divider and one multiplier. The iterative method can be easily scaled for higher accuracy (i.e. higher number of fractions) as it is quite suitable for pipelined implementation.

For non-iterative method, evaluation of polynomials of degree 7 and 6 is required for numerator and denominator respectively. For faster implementation, the numerator and denominator polynomials can be evaluated in parallel. Each degree requires one adder and one multiplier as given in (16). Apart from this, one divider is required in the last step which can be implemented using Newton Raphson method described above.

### 4.7 Tolerance to precision and input range

The accuracy varies with the precision and the input range. Depending on application, we can either chose a solution which provides better results across different precisions and ranges or a low-cost solution optimized for a given range and precision. Following table shows the parameters for different approximations for a maximum error of 1ulp. The alphabets used for approximations in the heading row correspond to those presented in table 1.

**Table 3.** Effect of input range and precision on approximation parameters

| Input Data | Output Data | Range | A | B1 | B2 | C | D | E |
|---|---|---|---|---|---|---|---|---|
| **S2.13** | S2.13 | ± 4 | 1/128 | 1/32 | 1/16 | 1/16 | 1/128 | 6 |
| **S2.13** | S.15 | ±4 | 1/128 | 1/32 | 1/16 | 1/64 | 1/256 | 6 |
| **S3.12** | S.15 | ±6 | 1/128 | 1/32 | 1/16 | 1/64 | 1/256 | 8 |
| **S2.5** | S.7 | ±4 | 1/8 | 1/32 | 1/32 | 1/8 | 1/8 | 4 |

### 4.8 Design Implementation Assessment

PWL is the simplest implementation but requires huge LUTs; so, can't be scaled easily. Quadratic Taylor series approximation provides as good results as a cubic Taylor series approximation or Catmull-Rom spline interpolation and should be preferred choice for implementation with medium accuracy and complexity. It was observed that the circuit runs faster if LUTs are used instead of combinatorial circuits for the interpolation vector in Catmull-Rom implementation or the derivatives (coefficients) for Taylor series expansion. However, the area is larger in this case.

Lambert's continuous function can be scaled for better accuracy compared to other approximations. It also has a nice pipelined implementation. However, it requires larger multipliers and a divider. The rational algorithms are better suited for pipelined implementations and the area and latency is more than the polynomial implementation. However, rational algorithms can be used for improving accuracy at a smaller incremental cost. Velocity factor is more adaptive to range post implementation; so, if the range is dynamically selected, then this is the best choice. The Lambert's continuous fraction is least adaptive.

For reasonable accuracy, the polynomial approximation such as PWL and Taylor series expansion yield good results. For more accurate implementation rational implementations have less incremental cost. The main drawback of rational approximations is the higher latency in computing tanh. However, if many back-to-back computations required in an application (e.g. neural network activations); then, the latency can be hidden for successive computations and throughput can be improved.

## 5 Conclusion

Multiple polynomial and rational approximation methods for fixed point tanh approximation and their hardware implementation are analyzed for accuracy and hardware implementation in this paper. Instead of choosing more complex implementation, one of these can be used for acceptable approximation. Though, this paper is presented in the context of deep learning accelerators, it can be easily adapted to other applications. The methods and implementation presented here will help the design community in selecting the right architecture for an application.